\documentstyle[prl,twocolumn,aps]{revtex}
\input{epsf}
\begin{document}
\draft
\twocolumn[\hsize\textwidth\columnwidth\hsize\csname @twocolumnfalse\endcsname
\title{Quantum Chaos Border for Quantum Computing}

\author{B. Georgeot and D. L. Shepelyansky}

\address {Laboratoire de Physique Quantique, UMR 5626 du CNRS, 
Universit\'e Paul Sabatier, F-31062 Toulouse Cedex 4, France}

\date{October 19, 1999; revised January 17, 2000}

\maketitle

\begin{abstract}
We study a generic model of quantum computer, 
composed  of many qubits coupled by short-range interaction.
Above a critical interqubit coupling strength, quantum chaos sets in, 
leading to quantum ergodicity 
of the computer eigenstates.  In this regime the noninteracting
qubit structure disappears, the eigenstates become complex and the operability
of the computer is destroyed. Despite the fact that the spacing between
multi-qubit states drops exponentially with the number of qubits $n$, we show
that the quantum chaos border decreases only linearly with $n$.  This
opens a broad parameter region where the efficient operation of
a quantum computer remains possible.
\end{abstract}
\pacs{PACS numbers: 03.67.Lx, 05.45.Mt, 24.10.Cn}
\vskip1pc]

\narrowtext
Since the pioneering work of Feynman \cite{feynman} and modern developments 
of efficient algorithms \cite{shor1}
and error-correcting codes \cite{shor2,steane1}, the realization of
quantum computers became a challenge of modern physics \cite{steane2}.  
Different experimental realizations have been proposed,
including ion traps \cite{zoller}, nuclear magnetic resonance systems
\cite{nmr}, nuclear spins with interaction controlled electronically
\cite{vagner,kane}, quantum dots \cite{loss}, Cooper pair boxes
\cite{cooper} and optical lattices \cite{lattice}.  
A key common feature of these experimental settings 
is the presence of interacting qubits (two-level systems).  Here we analyze
the effect of qubit interaction on operability of the quantum computer.
The interaction is required since a quantum computer needs to perform two-qubit
logical operation such as XOR \cite{steane2}.  We note that such a two-qubit
gate has been experimentally realized \cite{monroe}.

In  an isolated system of $n$ uncoupled qubits, 
the dimension of the total Hilbert space $N_H$
increases exponentially with $n$ ($N_H = 2^n$), while all eigenvalues 
of the Hamiltonian are included 
in an energy interval of size $\Delta E \sim n \Delta_0$, where $\Delta_0$
is the average energy distance between the two states of one qubit.
As a result, the average spacing $\Delta_n$ between
adjacent energy levels of the Hamiltonian decreases exponentially
with the number of qubits ($\Delta_n \sim n \Delta_0/N_H \ll \Delta_0$).  
When a coupling $J$ between the qubits is added ($J < \Delta_0$),
one still has $\Delta E \sim n \Delta_0$, $N_H$ is unchanged,
and the above estimate for
$\Delta_n$ still holds.
This general result for $\Delta_n$ is related to the 
exponentially large size $N_H$ of the Hilbert space, which
 is one of the main reasons of the striking 
efficiency of quantum computing \cite{feynman,shor1}.  
It implies that dense highly excited states
are needed for the computation. However, when performing the computation
one wants to operate with noninteracting multi-qubit states $|\psi_i>
=|\alpha_1,...,\alpha_n>$ where $\alpha_k=0,1$ marks the polarization
of each individual qubit.
These quantum register states should remain well-defined in the presence 
of interqubit coupling
even if multi-qubit levels are exponentially dense. Therefore the mixing of 
noninteracting multi-qubit states induced by the interaction is crucial for 
the computer operability.  In the field of quantum chaos \cite{houches,guhr}
it is known 
that noninteracting states will be eventually mixed by the interaction
and quantum ergodicity will set in: each quantum computer eigenstate 
will be composed of a large number
of noninteracting multi-qubit states $|\psi_i>$ and the original
quantum register states
will be washed out.
At first glance
one would expect that such mixing happens when the coupling between qubits
becomes comparable to the multi-qubit spacing $\Delta_n$.  In such a case,
the creation of quantum computers competitive with classical ones
would be rather difficult: since hundreds of qubits are necessary, this would
lead to absurdly strict restrictions on coupling strength.  Indeed, for
$n=1000$, the minimum number of qubits for which Shor's algorithm 
becomes useful \cite{steane2},  the multi-qubit spacing becomes
$\Delta_n \sim 10^3 \times 2^{-10^3} \Delta_0 \sim 10^{-298}$ K, 
where we used $\Delta_0 \sim 1$ K
that corresponds to the typical one-qubit spacing in the experimental
proposals \cite{vagner,kane}.   It is clear that the residual interaction $J$
between qubits in any experimental realization of the quantum computer 
will be larger than this.  For example, in the proposal \cite{kane},
the increase of effective electron mass by a factor of two, induced by the 
electrostatic
gate potential, means that the spin-spin interaction is changed
from $J \sim \Delta_0 \sim 1$ K (corresponding to a distance
between donors of $200 $ {\AA} and an effective 
Bohr radius of $30$ {\AA} in Eq.2 of \cite{kane}) 
to the residual interaction $J \sim 10^{-5}$ K $\gg \Delta_n$.

However the problem is not so simple, since the interaction 
is always of two-body nature and not all of the multi-qubit states
are directly coupled.  Actually the number of states directly coupled
to such a quantum register state $|\psi_i>$ 
increases not faster than quadratically with 
$n$.  A similar problem appears in other physical many-body interacting systems 
such as nuclei, complex atoms,
quantum dots and quantum spin glasses
\cite{french,aberg,zelevinsky,sivan,1997}.
It was realized that 
sufficiently strong interaction leads to quantum chaos and
internal (dynamical) thermalization,  
where the eigenstates properties follow the predictions of Random Matrix Theory
(RMT) \cite{houches,guhr,french,aberg,zelevinsky}.  
The quantum
chaos border for this dynamical thermalization has been established only 
recently and it has been shown that the relevant coupling strength should
be larger than the energy spacing between directly coupled states $\Delta_c$
\cite{aberg,1997}.
Since $\Delta_c$ drops algebraically with $n$, it is exponentially larger than 
$\Delta_n \sim n 2^{-n}\Delta_0$, and therefore a relatively
large coupling strength is required for the emergence of quantum chaos 
and ergodicity.
A similar border for interacting qubit systems would allow a reasonable
regime of operability for quantum computers.

To investigate the emergence of quantum chaos in quantum computers,
we chose a model of $n$ qubits on a two-dimensional lattice
with nearest-neighbour interqubit coupling.  The Hamiltonian reads:
\begin{equation}
\label{hamil}
H = \sum_{i} \Gamma_i \sigma_{i}^z + \sum_{i<j} J_{ij} 
\sigma_{i}^x \sigma_{j}^x,
\end{equation}
where the $\sigma_{i}$ are the Pauli matrices for the qubit $i$ and the second
sum runs
over nearest-neighbour qubit pairs with periodic boundary conditions applied.
The energy spacing between the two states of a qubit is represented 
by $\Gamma_i$ randomly and uniformly distributed in the interval 
$[\Delta_0 -\delta /2, \Delta_0 + \delta /2 ]$.  The parameter $\delta$
gives the width of the distribution near the average value $\Delta_0$ 
and varies from $0$ to $\Delta_0$.
Here $\Gamma_i$ can be viewed as the splitting
of nuclear spin levels in a local magnetic field, as it is discussed 
in the experimental
proposals \cite{vagner,kane}.  The different values of $\Gamma_i$ 
are needed to prepare a specific initial state by electromagnetic pulses
in nuclear magnetic resonance.  
In this case the couplings $J_{ij}$ will
represent the hyperfine interaction between the spins, which is needed
to build the quantum computer.   Different physical mechanisms can
generate these couplings, such as spin exciton exchange \cite{vagner,kane}, 
dipole-dipole interaction, etc... For generality we chose 
 $J_{ij}$ randomly distributed in the interval $[-J,J]$.  The Hamiltonian 
(\ref{hamil}) can be considered as a generic quantum computer model, which
catches the main physics of different experimental proposals.  For example 
a similar Hamiltonian appears in a quantum computer based on optical lattices
\cite{lattice,molmer}.
We restrict 
ourselves to the case of static couplings which are always present as a
residual interaction and are much larger than the multi-qubit spacing
$\Delta_n$ even for moderate values of $n$. 
In a sense (\ref{hamil}) describes the hardware of the computer,
while gates operation in time requires additional studies, which are possible
only if the properties of the hardware are well understood.

As is well known in the field of quantum chaos, the transition to ergodic
eigenstates is reflected in the level spacing statistics $P(s)$, which
goes from the Poisson distribution $P_P(s)=\exp(-s)$ for nonergodic states
to the Wigner-Dyson (WD) distribution $P_W(s) = (\pi s/2)\exp(-\pi s^2/4)$,
corresponding to RMT, for ergodic states.  
Here $s$ is the nearest level spacing
measured in units of average spacing and $P(s)$ is the probability
to find two adjacent levels whose spacing is in $[s,s+ds]$.
\begin{figure}
\epsfxsize=3.4in
\epsfysize=2.6in
\epsffile{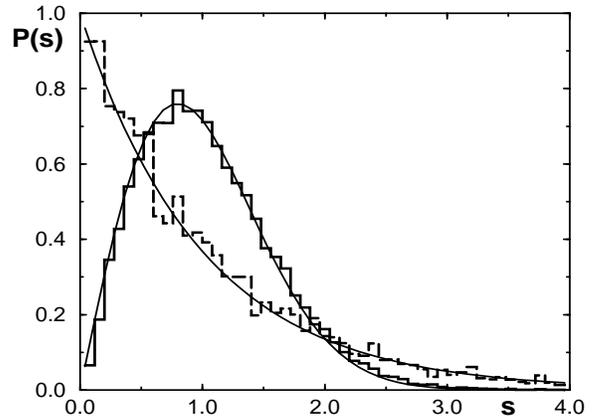}
\vglue -0.2cm
\caption{Transition from Poisson to WD statistics in the model 
(\ref{hamil}) for the states
in the middle of the energy band ($\pm 6.25 \% $ around the center) for
$n$=12 : $J/\Delta_0=0.02, \eta=1.003$ (dashed line histogram);
$J/\Delta_0=0.48, \eta=0.049$ (full line histogram). Full curves
 show $P_P(s)$ and $P_W (s)$; $N_S > 2.5 \times 10^4$,
 $N_D=100$, $\delta=\Delta_0$.} 
\label{fig1}
\end{figure}

\begin{figure}
\epsfxsize=3.4in
\epsfysize=2.6in
\epsffile{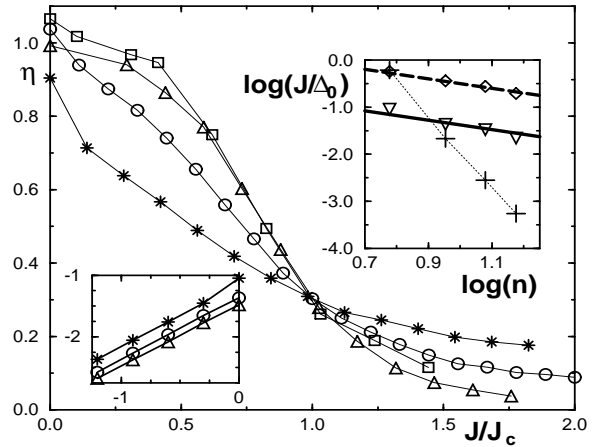}
\vglue -0.2cm
\caption{Dependence of $\eta$ on the rescaled coupling strength $J/J_{c}$
for the states in the middle of the energy band  for
$n=6 (*),9 $(o)$,12 $(triangles)$,15$(squares); 
$\delta=\Delta_0$.
The upper insert shows $\log(J_c/\Delta_0)$ (diamonds) 
and $\log(J_{cs}/\Delta_0)$ (triangles) 
versus $\log(n)$; the variation of the scaled multi-qubit
spacing $\Delta_n/\Delta_0$ with $\log(n)$ is shown for comparison (+).
Dashed line gives the theoretical formula 
$J_c=C \Delta_0/n$ with $C=3.16$; the solid line is $J_{cs}=0.41 \Delta_0/n$.
The lower insert shows $\log(J_{cs}/\Delta_0)$ versus $\log(\delta/\Delta_0)$
for $n=6 (*),9$ (o), 12 (triangles); straight lines have
slope $1$.}
\label{fig2}
\end{figure}
\noindent The majority of our data are displayed for 
the middle of the energy spectrum,
where the transition starts, and which therefore sets the limit of
operability of the quantum computer.  The model (\ref{hamil}) has 
two symmetry classes characterized by an odd or even number
of qubits up, and the data are given for one symmetry class.
In order to reduce statistical fluctuations, 
we use $5 \leq N_D \leq 4 \times 10^4$ random realizations of $\Gamma_i$
and $J_{ij}$, as is done usually in RMT \cite{guhr}.  Eigenvalues and
eigenvectors are computed by exact diagonalization of the Hamiltonian
matrix (\ref{hamil}) for each realization.
In this way the total number of
spacings is $10^4 < N_S \leq 1.6 \times 10^5$ ($N_S \propto N_D N_H$).
An example of the transition in the spectral statistics is shown in Fig.1.

To analyze the evolution of $P(s)$ with the
coupling $J$, it is convenient to use the parameter 
$\eta=\int_0^{s_0}
(P(s)-P_{W}(s)) ds / \int_0^{s_0} (P_{P}(s)-P_{W}(s)) ds$,
where  $s_0=0.4729...$ is the intersection point of $P_P(s)$ and $P_{W}(s)$.
In this way $P_P(s)$
corresponds to $\eta=1$, and $P_W(s)$ to $\eta=0$. As is usual in the field of
quantum chaos, the variation of $\eta$ characterizes the evolution of $P(s)$
\cite{1997}.
The variation of $\eta$ with respect to $J/\Delta_0$ is presented in Fig.2
for $\delta=\Delta_0$
showing that indeed $\eta$ drops from $1$ to $0$ with increasing coupling
strength.  The transition appears to become sharper for larger system sizes.
The typical $J_c$ value near which the transition takes place corresponds
to intermediate values of $\eta$.  We chose the condition $\eta(J_c)=0.3$.
The dependence of $J_c$ on $n$ is given in the
Fig.2.  In analogy with other many-body systems discussed 
in \cite{aberg,1997}, we expect
that $J_c \approx \Delta_c \approx C \Delta_0/n$, where $C$ is some
numerical constant.  Indeed, one multi-qubit state is coupled
to $2n$ other states 
in an energy interval of order $6\Delta_0$.  This theoretical
estimate is in agreement with the data of Fig.2, with $C \approx 3$.
We stress that this critical coupling is exponentially larger than the
multi-qubit level spacing $\Delta_n \sim n 2^{-n} \Delta_0$,
as is shown on Fig.2.  For the case $\delta \ll  \Delta_0$,
the total spectrum at $J=0$ is composed of $n$ bands with interband distance
$2\Delta_0$ and a bandwidth of $\sqrt{n} \delta$.  Within one band, one
multi-qubit state is coupled to about $n$ states in an energy interval
of $2 \delta $, so that $J_c \approx \Delta_c \sim \delta /n$.  This 
quantum chaos border is still much bigger than 
$\Delta_n \sim \sqrt{n}\delta / (N_H/n) \sim n^{3/2} 2^{-n} \delta$.  

The transition in the level statistics reflects the drastic change
in the multi-qubit structure of the eigenstates of (\ref{hamil}).
Indeed, Fig.3 shows that for $J < J_c$ one eigenstate is formed
only by one or few noninteracting states $|\psi_i>$, while for
$J>J_c$ a huge number of them are required.  In the latter case, 
the computer eigenstates become a random mixture of 
quantum register states $|\psi_i>$,
making rather difficult to perform computation.

To study this drastic change in the structure of eigenstates, it
is convenient to use the quantum eigenstate entropy $S_q$, defined by:
$S_q = -\sum_{i} W_i \log_2 W_i$, where $W_i$ is the quantum probability
to find the noninteracting multi-qubit state $|\psi_i>$ in the eigenstate 
$|\phi>$ of (\ref{hamil}) ($W_i=|<\psi_i|\phi>|^2$).  In this way $S_q=0$
if $|\phi>$ is one noninteracting state ($J=0$), 
$S_q=1$ if $|\phi>$ is equally
composed of two $|\psi_i>$, and the maximal value is $S_q=n$ if all
$2^n$ states contribute equally to $|\phi>$.  The variation of the 
average quantum entropy with $J$ is shown on Fig.4
for $\delta=\Delta_0$.
It shows that $S_q$ grows with $J$ and the transition to ergodic states with
large $S_q$ takes place in the vicinity of $J_c$.   
In addition these 
data show that the critical coupling $J_{cs}$ at which $S_q=1$ is
$J_{cs} \approx 0.13 J_c$.  The ratio $J_{cs}/J_c$ stays within $15 \%$
of the average value  when $n$ changes from $6$ to $15$, while 
the ratio $\Delta_n/J_c$ varies
from $1$ to $3\times 10^{-3}$  (see upper insert of Fig.2).  The dependence
of $J_{cs}$ on $\delta$ is shown on the lower insert of Fig.2; it clearly shows
the linear decrease of $J_{cs}$ with $\delta$ and can be well described
by $J_{cs}= 0.4 \delta/n$.  Naturally, the quantum chaos border drops 
to zero with $\delta$ due to the quasidegeneracy inside 
the energy bands at $J=0$.

\begin{figure}
\epsfxsize=3.4in
\epsfysize=2.6in
\epsffile{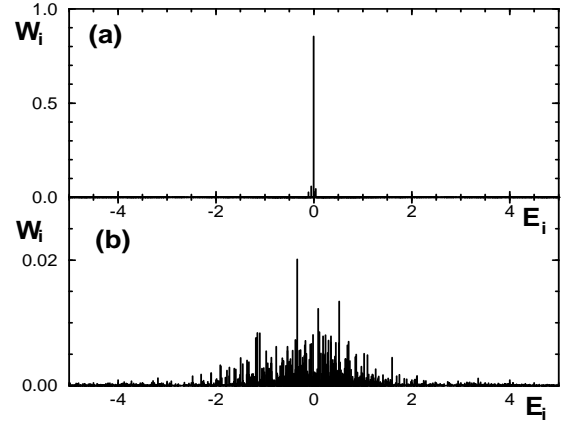}
\vglue -0.2cm
\caption{Two quantum computer
eigenstates of model (\ref{hamil}) in the basis of noninteracting 
multi-qubit states, i.e. $W_i=|<\psi_i|\phi>|^2$ as a function of 
noninteracting multi-qubit energy
$E_i$ for $n=12$  and $\delta=\Delta_0$
with $J_c/\Delta_0=0.273$ (see text):  
(a) $J/\Delta_0=0.02$; (b) $J/\Delta_0=0.48$.}
\label{fig3}
\end{figure}

\vskip -0.7cm
\begin{figure}
\epsfxsize=3.4in
\epsfysize=2.6in
\epsffile{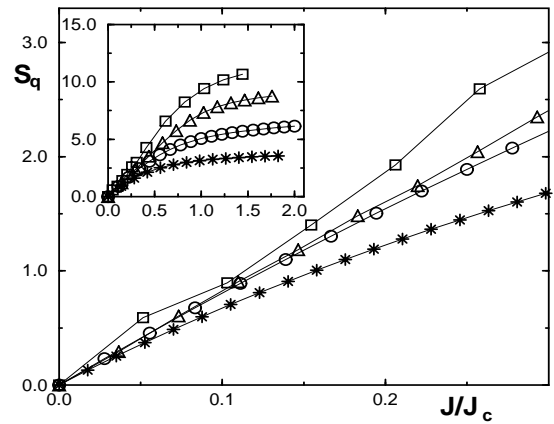}
\vglue -0.2cm
\caption{Dependence of the quantum eigenstate entropy $S_q$ on $J/J_c$ for
$\delta=\Delta_0$ and $n=6 (*)$, 9  (o), 12 (triangles), 15 (squares); 
$10^4 < N_S \leq 1.6 \times 10^5$. 
Insert shows the dependence on larger scale.} 
\label{fig4}
\end{figure}

We note that for
$n=1000$ and $\delta=\Delta_0=1$ K, only two multi-qubit states will be mixed
at $J_{cs} \approx 0.4 \Delta_0 /n \approx 0.4 $ mK.  This critical coupling
is much larger than the multi-qubit level spacing $\Delta_n \sim 10^{-298}$ K.
Even if the quantum border $J_{cs}$ corresponds to a relatively low 
coupling strength it seems reasonable that the residual interaction between
qubits can stay below this threshold with current technologies (but not below 
$\Delta_n$).

\begin{figure}
\epsfxsize=3.0in
\epsfysize=3.0in
\epsffile{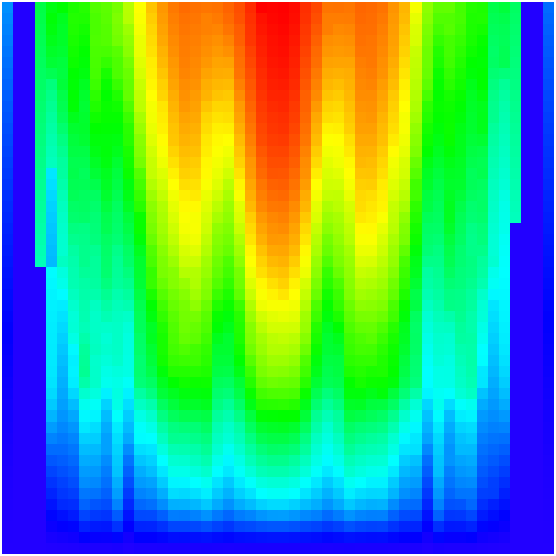}
\vglue 0.2cm
\caption{The quantum computer melting induced by the coupling between qubits.
Color represents the level of quantum eigenstate entropy $S_q$, from bright red
($S_q \approx 11$) to blue
($S_q =0$).  Horizontal axis is the energy of the computer eigenstates 
counted
from the ground state to the maximal energy ($\approx 2n\Delta_0$).
Vertical axis is the value of $J/\Delta_0$, varying from $0$ to $0.5$. Here
$n=12$, $\delta=\Delta_0$, $J_c/\Delta_0=0.273$, and
one random realization of (\ref{hamil}) is chosen.} 
\label{fig5}
\end{figure}

The pictorial image of the quantum computer melting under the influence
of the interqubit coupling $J$ is shown on Fig.5.  The melting starts 
in the middle of the spectrum (high energy) and progressively invades
low-energy states and the whole computer, destroying its operability.
We stress that this destruction takes place in an isolated
system without any external decoherence process.  Nevertheless the
thermalization in this closed
system, which appears because of the interqubit coupling, can mimic
the effect of a coupling with the external world and external decoherence.
Above the quantum chaos border an initial register
state $|\psi_i>$ will spread quickly with time \cite{note}
over an exponential 
number of eigenstates of the system with residual interaction, destroying
gates operability. 

Our studies show that the mixing of multi-qubit states and onset of quantum
chaos induced by interqubit coupling leads to the melting of a realistic
quantum computer and destruction of its operability; however, the quantum
chaos border found for this process corresponds to a relatively strong
interaction, being exponentially larger than the energy level
spacing between multi-qubit
states.  We expect that below this border, error-correcting codes 
\cite{shor2,steane1} will 
allow to perform efficient quantum computing with a large number of qubits.
However we note that quantum chaos sets in very easily if the fluctuation 
amplitude $\delta$ of individual qubit spacing drops to zero ($J_c \propto
\delta$).

We thank O.P. Sushkov and I.D. Vagner for stimulating
discussions, and the IDRIS in Orsay and the CICT in Toulouse for access to 
their supercomputers.


\vskip -0.5cm
\begin{thebibliography}{99}
\bibitem{feynman} R.~P.~Feynman,
Found. Phys. {\bf 16}, 507 (1986).
\bibitem{shor1} P.~W.~Shor, in Proc. 35th Annu. Symp. Foundations of
Computer Science (ed. Goldwasser, S. ), 124 (IEEE Computer Society, Los
Alamitos, CA, 1994).
\bibitem{shor2} A.~R.~Calderbank  and P.~W.~Shor, 
Phys. Rev. A {\bf 54}, 1098 (1996).
\bibitem{steane1} A.~Steane,
Proc. Roy. Soc. Lond. A {\bf 452}, 2551 (1996).
\bibitem{steane2}  A. Steane, Rep. Progr. Phys. {\bf 61},
117 (1998).
\bibitem{zoller} J.~I.~Cirac and P.~Zoller,
Phys. Rev. Lett. {\bf 74}, 4091 (1995).
\bibitem{nmr} N.~A.~Gershenfeld and I.~L.~Chuang, 
 Science {\bf 275}, 350 (1997); D.~G.~Cory, A.~F.~Fahmy and T.~F.~Havel,
In Proc. of the 4th Workshop on Physics and Computation
(Complex Systems Institute, Boston, MA, 1996).
\bibitem{vagner} V.~Privman, I.~D.~Vagner and G.~Kventsel, 
Phys. Lett. A {\bf 239}, 141 (1998).
\bibitem{kane} B.~E.~Kane, Nature {\bf 393}, 133 (1998).
\bibitem{loss} D.~Loss  and D.~P.~Di~Vincenzo,
 Phys. Rev. A {\bf 57}, 120 (1998).
\bibitem{cooper} Y.~Nakamura, Yu.~A.~Pashkin, and J.~S.~Tsai,  
Nature {\bf 398}, 786 (1999).
\bibitem{lattice} G.~K.~Brennen, C.~M.~Caves, P.~S.~Jessen and I.~H.~Deutsch
Phys. Rev. Lett. {\bf 82}, 1060 (1999); D.~Jaksch, H.~J.~Briegel, 
J.~I.~Cirac, C.~W.~Gardiner and P.~Zoller,
 Phys. Rev. Lett. {\bf 82}, 1975 (1999).
\bibitem{monroe} C.~Monroe, D.~M.~Meekhof, B.~E.~King, W.~M.~Itano
and  D.~J.~Wineland, Phys. Rev. Lett. {\bf 75}, 4714 (1995).
\bibitem{houches} {\it Les Houches Lecture Series} {\bf 52},
        Eds. Giannoni, M.-J., Voros, A. and Zinn-Justin, J. (North-Holland,
        Amsterdam, 1991).
\bibitem{guhr} T. Guhr, A. M\"uller-Groeling and H.~A.~Weidenm\"uller, 
Phys. Rep. {\bf 299}, 189 (1999).
\bibitem{french} J.~B.~French and S.~S.~M.~Wong, 
Phys. Lett. B {\bf 33}, 449 (1970); O.~Bohigas and J.~Flores, 
 {\em ibid} {\bf 34}, 261 (1971).
\bibitem{aberg} S. {\AA}berg, Phys. Rev. Lett. {\bf 64}, 3119 (1990).
\bibitem{zelevinsky} V.~Zelevinsky, B.~A.~Brown, N.~Frazier and M.~Horoi, 
 Phys. Rep. {\bf 276}, 85 (1996); V.~V.~Flambaum, F.~M. Izrailev, and G.~Casati, 
Phys. Rev. E {\bf 54}, 2136 (1996).
\bibitem{sivan} U.~Sivan,  F.~P.~Milliken, K.~Milkove, S.~Rishton,
Y.~Lee, J.~M.~Hong, V.~Boegli, D.~Kern, and M.~de~Franza,
Europhys. Lett. {\bf 25}, 605 (1994).
\bibitem{1997} D.~L.~Shepelyansky and O.~P.~Sushkov,
Europhys. Lett. {\bf 37}, 121 (1997); P.~Jacquod and  D.~L.~Shepelyansky, 
Phys. Rev. Lett. {\bf 79}, 1837 (1997); A.~D.~Mirlin and Y.~V.~Fyodorov,
Phys. Rev. B {\bf 56}, 13393 (1997); D.~Weinmann, J.-L.~Pichard and Y.~Imry,
J. Phys. I France
{\bf 7}, 1559 (1997); B.~Georgeot and D.~L.~Shepelyansky,
Phys. Rev. Lett. {\bf 81},
5129 (1998).
\bibitem{molmer} A.~S\o rensen and K.~M\o lmer,
Phys. Rev. Lett. {\bf 83}, 2274 (1999).
\bibitem{note} As in many-body systems \cite{1997}
the spreading rate $\gamma$ can be estimated as  
$\gamma \sim J^2/\Delta_c \sim n J^2/\delta$
for $J_c < J < \delta$.
\end{thebibliography}
\end{document}